# Beyond The Fermi's Golden Rule: Discrete-Time Decoherence Of Quantum Mesoscopic Devices Due To Bandlimited Quantum Noise

Evgeny A. Polyakov

We are in the midst of second quantum revolution where the mesoscopic quantum devices are actively employed for technological purposes. Despite this fact, the description of their real-time dynamics beyond the Fermi's golden rule remains a formiddable theoretical problem. This is due to the rapid spread of entanglement within the degrees of freedom of the surrounding environment. This is accompanied with a quantum noise (QN) acting on the mesoscopic device. In this work we propose a possible way out: to exploit the fact that this QN is usually bandlimited. This is because its spectral density is often contained in peaks of localized modes and resonances, and may be constrained by bandgaps. Inspired by the Kotelnikov sampling theorem from the theory of classical bandlimited signals, we put forward and explore the idea that when the QN spectral density has effective bandwidth $B$, the quantum noise becomes a discrete-time process, with an elementary time step $\tau \propto B^{-1}$. After each time step $\tau$, one new QN degree of freedom (DoF) gets coupled to the device for the first time, and one new QN DoF get irreversibly decoupled. Only a bounded number of QN DoFs are significantly coupled at any time moment. We call these DoFs the *Kotelnikov modes*. As a result, the real-time dissipative quantum motion has a natural structure of a discrete-time matrix product state, with a bounded bond dimension. This yields a microscopically derived collision model. The temporal entanglement entropy appears to be bounded (area-law scaling) in the frame of Kotelnikov modes. The irreversibly decoupled modes can be traced out as soon as they occur during the real-time evolution. This leads to a novel *bandlimited* input-output formalism and to quantum jump Monte Carlo simulation techniques for real-time motion of open quantum systems. We illustrate this idea on a spin-boson model.

## I. INTRODUCTION

The last decade has witnessed a rapid develoment of quantum technology [1]. This technology is concerned with measurement, control, communication, and computation based on quantum mesoscopic degrees of freedom. The motivation behind this development is two-fold. The first is miniaturization: the technological devices on a nanometer scale must be designed according to the laws of quantum physics. The second one is the promise to achieve a significant supremacy over the technology based on the laws of classical physics [1]. The role of the quantum mesoscopic degrees of freedom can be played by e.g. selected electron levels of trapped ions [2], semiconductor quantum dots [3], superconducting circuits [4], spin of electrons in nitrogen-vacancy centers [5], Majorana zero modes [6]. In any physical device, these dedicated degrees of freedom are always coupled to an infinite number of other degrees of freedom. The latter are called the environment. This way we arrive to the model of open quantum system (OQS) [7], which is thus a topic of active research. Part of the environment can be used to control the open quantum system (e.g. transmission line resonator in superconducting circuits [8, 9]). The other part is responsible for irreversible scattering of the quantum information (decoherence) [7, 9]. The latter is the major obstacle to the creation of scalable quantum computers [9], which stimulates the research on decoherence phenomena.

In the field of condensed matter the open quantum system is referred to as the quatum impurity model (QIM), especially in the solid state context. The role of QIM may be played by a single lattice site, localized spin, or some other defect. QIM provides a model of solid state mesoscopic quantum devices [3]. Besides that, at low temperatures QIM develops complicated many-body correlations with the itinerant electrons. The latter play the role of the environment, and QIM becomes an open quantum system. These correlations significantly affect the transport properties of materials [10] and properties of quantum devices [11]. Currently there is an active investigation of non-equilibrium properties of QIM [12, 13], in particular under periodical driving [14].

Finally, besides the technological importance, the model of open quantum system plays important role in the microscopic foundations of non-equilibrium thermodynamics [15–17]: how the time arrow emerges from the reversible microscopic dynamics [18, 19]; statistics of work, heat, and entropy production [16, 17, 20–23]; how the thermodynamics is related to quantum trajectories [24–27]; the studies of quantum heat engines [28–30].

All of the above shows that at the moment the actual problem is to study the non-stationary real-time motion of OQS/QIM, in order to clarify open questions in the research fields mentioned above.

The description of real-time motion of OQS/QIM is clear and simple when we can adopt the Fermi's golden rule (the so-called Markovian approximation) [7]: OQS experiences sudden transitions (quantum jumps) $\omega_1 \to \omega_2$ between the energy eigenstates $|\omega_1\rangle$ and $|\omega_2\rangle$ with a rate $\Gamma_{1 \to 2} = \left| \left\langle \omega_2 \left| \widehat{V} \right| \omega_1 \right\rangle \right|^2 J(\omega_1 - \omega_2) 2\pi/\hbar$. Here $\widehat{V}$ is coupling to the environment, and $J(\omega_1 - \omega_2)$ is the effective density of degrees of freedom (DoFs) of the environment seen by OQS. The Markovian approximation is valid when the density of states $J(\omega)$ is smooth on the scales of OQS lineshapes. The resulting stochastic dynamics is described by the so-called Lindblad master equation [7]. Due to such a clear probabilistic picture, this Markovian approximation is well understood nowadays. It is a workhorse in describing the quantum optics experiments [31] and the physical mechanisms of decoherence in qubits [32]. There exist efficient numerical simulation techniques [7]. The Markovian approximation is also used in the simulations of quantum heat engines [33] and in foundations of non-equilibrium quantum thermodynamics [34]. The Fermi's golden rule corresponds to a white (delta-correlated in time) quantum noise [35, 36].

The real-time motion of OQS becomes a formiddable theoretical problem outside the applicability of the Fermi's golden rule (the so-called non-Markovian regime). This happens when the spectral density $J(\omega)$ of the environment is not flat, but on the contrary, is structured on the scales of OQS lineshapes. The structured $J(\omega)$ naturally occurs in quantum devices which are based on solid state impurities. For example, in the nitrogen-vacancy centers in the diamond [37], the impurity (defect) is coupled to the environment of acoustic phonons. The corresponding $J(\omega)$ develops a number of peaks corresponding to vibrational resonances and localized modes. Analogous situation occurs for impurities in semiconductors [38]. The structured $J(\omega)$ also occurs in the photonic band gap materials [39] where $J(\omega)$ is organized into allowed and forbidden ($J(\omega) \approx 0$) bands. Finally, there is experimental progress in creating artificial environments with a structured band-gapped $J(\omega)$ [40]. The motivation for the latter is to introduce effects into the motion of OQS which are beyond the Markovian approximation e.g. the information backflow [41]. One hopes to exploit such non-Markovian effects as a resource for the quantum technologies (the so-called reservoir enegineering) [42].

The deep reason behind the difficulty of non-Markovian regime is the apparent absence of the full-fledged quantum jumps. Indeed, the Fermi's golden rule provides us with quantum jumps which are completed and irreversible elements of classical reality. As a result, they have a probabilistic interpretation, which leads to a beautiful formal picture (the divisibility of quantum Markovian motion and the concept of dynamical semigroups) [7], and to efficient Monte-Carlo simulation methods [7]. On the contrary, beyound the Markovian approximation, every excitation which is emitted by OQS can always be reabsorbed back (return to OQS) after a time interval $t$ with some amplitude $C_q(t)$. This amplitude slowly decreases as a certain inverse power $C_q(t) \propto t^{-p}, p > 0$ [43–45]. As a result, one fails to introduce the concept of full-fledged non-Markovian quantum jumps [46, 47]: they appear to be never complete in finite time. For any finite time, there is a non-nebligible amplitude that the excitation will return. Therefore, the entanglement rapidly propagates through the surrounding environment, with a combinatorial growth of the complexity of real-time quantum motion (the so-called temporal entanglement barrier [44, 48–50]).

In literature there are many attemps to solve the puzzle of non-Markovian real-time motion. Nowadays the most activity is concentrated on combining the Feynman-Vernon influence functional [51] with the tensor network techniques [14, 48, 50, 52–55]. At the heart of these attemps is a bruteforce numerical compression of many-body quantum states using smart linear algebra machinery and the singular value decomposition (SVD) of high-rank tensors.

In this work we continue another line of research: our aim is to find a clear visual physical principle, which could replace/extend the Fermi's golden rule in the non-Markovian regime. We started in [45], where we have employed the time-domain wavefunction to identify how OQS/QIM forgets about the previously emitted quanta. We have found that all the emitted spectral content is forgotten except progressively small vicinities of frequencies $\omega_k$ where $J(\omega_k)$ has sharp features (e.g. band edges, sharp peaks). In [44] we have implemented this approach in a numerical method and conjectured that the time domain should be employed in a properly formulated real-time renormalization group (RTRG). Such RTRG was proposed in [56] which was recently published as [43]. The present paper further develops this approach by taking into account the spectral properties of the quantum noise.

This work is structured as follows. In sec. II we present the model of open quantum system and state the problem considered. Then in sec. III we introduce the concept of bandlimited quantum noise and bear ananlogies with the classical Kotelnikov sampling. In sec. IV we introduce the quantum analog of Kolelnikov sampling in the context of interaction quench. The resulting discrete-time model, simulation techniques, and physical picture are described. In sec. V we derive the discrete-time model and provide a numerical recipe to find its coupling constants. We conclude in VI. There are two appendices with some details.

## II. THE MODEL

In this work we study the real time motion of a small open quantum system with a Hamiltonian $\hat{H}_s$. It is surrounded by a quantum environment $\hat{H}_b$. We assume there is some local site in the environment which is represented by annihilation $\hat{a}_0$ and creation $\hat{a}_0^\dagger$ operators, $\left[\hat{a}_0, \hat{a}_0^\dagger\right]_\xi = 1$, with $\xi = \pm 1$ depending on the statistics of the environment's excitations (fermions or bosons). OQS is coupled to this site via some operators $\hat{V}, \hat{V}^\dagger$. The joint Hamiltonian for this model is

$$\hat{H} = \hat{H}_s + \hat{V}^\dagger \hat{a}_0 + \hat{V} \hat{a}_0^\dagger + \hat{H}_b. \quad (1)$$

Here for simplicity we assume the bilinear coupling between the OQS operators $\hat{V}, \hat{V}^\dagger$ and between the environment's $\hat{a}_0$ and $\hat{a}_0^\dagger$.

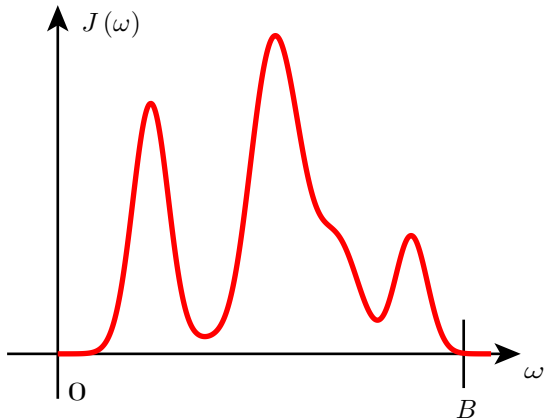

Figure 1. In this work we consider the quantum environment whose spectral density $J(\omega)$ is effectively bounded to some frequency range $[0, B]$, but otherwise can be arbitrary complex.

We study the joint quantum state $|\Psi(t)\rangle$ of OQS and environment. The state $|\Psi(t)\rangle$ may appear as a result of the interaction quench. That is, we assume that initially the environment is in its ground state $|0\rangle_{\rm b}$, and OQS is in some state $|\psi\rangle_{\rm s}$: $|\Psi(0)\rangle = |\psi\rangle_{\rm s} \otimes |0\rangle_{\rm b}$. Then at $t = 0$ we couple OQS to the environment and track the resulting evolution.

In this work we consider a specific form of the environment, a semi-infinite chain of sites

$$\widehat{H}_{\rm b} = \sum_{j=0}^{\infty} \left\{ \varepsilon_j \widehat{a}_j^\dagger \widehat{a}_j + h_j \widehat{a}_j^\dagger \widehat{a}_{j+1} + h_j \widehat{a}_{j+1}^\dagger \widehat{a}_j \right\}, \quad (2)$$

with $\left[ \widehat{a}_k, \widehat{a}_l^\dagger \right]_\xi = \delta_{kl}$. Here the coupling site operators $\widehat{a}_0$ and $\widehat{a}_0^\dagger$ appear as the first sites of the chain. This chain representation is general: given a spectral density $J(\omega)$, there is an algorithm how to compute $\varepsilon_j, h_j$ so that the resulting chain reproduces $J(\omega)$ [57–61] as seen by OQS.

## III. BANDLIMITED QUANTUM NOISE. ANALOGY TO THE CLASSICAL KOTELNIKOV SAMPLING THEOREM

A good physical principle requires good assumptions. In this work we assume that the spectral density $J(\omega)$ of the environment is effectively bounded to some frequency range $[0, B]$, see Fig. 1. In most cases (e.g., environment with non-interacting quasiparticles) this means that the quantum field emitted into the environment also has a bandlimited spectrum. From the discipline of classical signal processing we know the Kotelnikov sampling theorem [62], which is the following. Consider a random classical signal $z(t)$. For each time moment $t$ the value of the signal $z(t)$ is a complex number. Its time correlator is

$$C_{\rm cl}(t) = \overline{z^*(t)\, z(0)}, \quad (3)$$

where the averaging $\overline{[\cdot]}$ is over the signal samples. The spectral density $J_{\rm cl}(\omega)$ of the signal is provided by the causal Fourier transform of the correlator

$$J_{\rm cl}(\omega) = {\rm Re} \int_0^{+\infty} dt\, C_{\rm cl}(t)\, e^{i\omega t - 0t}. \quad (4)$$

Suppose that the signal turns out to have a finite band halfwidth $B$, see Fig. 2, a). Then the signal can be faithfully (without loss of information) represented by its values at discrete time moments $t_k = k\tau$, $\tau = 1/2B$, with integer $k$, Fig. 2 b):

$$z(t) = \sum_{k=-\infty}^{\infty} z(t_k) \frac{\sin(2\pi B(t - t_k))}{2\pi B(t - t_k)}. \quad (5)$$

The message of the Kotelnikov sampling is that in each finite time interval $T$ there is only a finite number $m(t) \propto TB$ of significant DoFs of the signal, see A. These DoFs are given by the sinc basis functions, $\kappa_k(t) = \sin(2\pi B(t - t_k))/2\pi B(t - t_k)$. Each DoF $\kappa_k$ is in one-to-one correspondence with its time moment $t_k = k\tau$.

Now we return to the quantum case. Analogously to the classical correlator eq. (3), we introduce the correlator of the zero-point quantum noise [43, 44, 63],

$$C_{\rm q}(t) = {}_{\rm b}\langle 0| \widehat{a}_0(t)\, \widehat{a}_0^\dagger(0)\, |0\rangle_{\rm b} \quad (6)$$

for the vacuum state $|0\rangle_{\rm b}$ of environment. Here we employ the partial interaction picture with respect to the free motion of environment, $\widehat{a}_0(t) = \exp\left( it\widehat{H}_{\rm b} \right) \widehat{a}_0 \exp\left( -it\widehat{H}_{\rm b} \right)$.

Analogously to the classical spectral density eq. (4), the spectral density of the zero-point quantum noise (which coincides with the spectral density of the environment) $J(\omega)$ is also a Fourier transform [43, 44]

$$J(\omega) = {\rm Re} \int_0^{+\infty} dt\, C_{\rm q}(t; 0)\, e^{i\omega t - 0t}. \quad (7)$$

If $J(\omega)$ is bandlimited, we expect that, analogously to the classical case, the quantum noise should be faithfully represented by a discrete time process.

## IV. KOTELNIKOV SAMPLING OF BANDLIMITED QUANTUM NOISE: THE RESULTING MODEL

In this section we present the resulting model of quantum dissipative motion under bandlimited quantum noise. It is ready for use in calculations of specific problems. We also discuss some physical implications. In the following sections we justify this model and present the algorithm how to calculate the numerical values of the model constants.

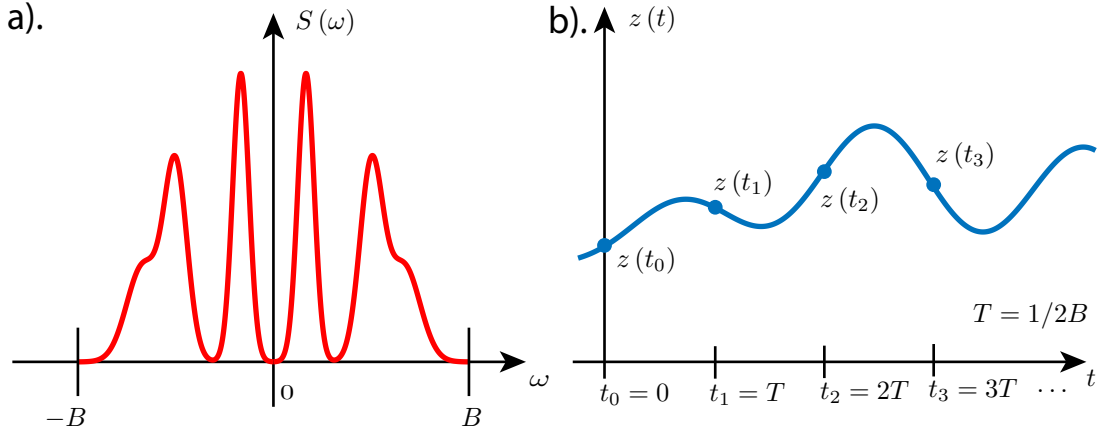

Figure 2. The classical Kotelnikov sampling theorem states that a) if a signal $z(t)$ has a spectral density $J_{\text{cl}}(\omega)$ which is limited to a frequency range $[-B, B]$, then b) the signal $z(t)$ is completely determined (without loss of information) by a sequence of its values $\ldots, z(t_{-1}), z(t_0), z(t_1), \ldots z(t_k) \ldots$, at discrete times moments which are spaced $\tau = 1/2B$ seconds apart.

### A. Forward quantum Kotelnikov sampling

Notice that the statement of the real-time problem is slightly different from that of the original classical Kotelnikov theorem eq. (5): we have the initial value problem instead of the signal on the whole time axis. Therefore, we need a causal variant of the Kotelnikov theorem, which we call the *forward Kotelnikov sampling*, see appendix A. According to the latter, we expect that the DoFs of the environment $\widehat{a}_0, \widehat{a}_1, \ldots$ can be arranged in a sequence, $\widehat{\kappa}_k^{\text{in}}$, $k = 1, 2, 3 \ldots$ by some change of frame (Bogoliubov transform) $W$,

$$\begin{bmatrix} \widehat{\kappa}_1^{\text{in}} \\ \widehat{\kappa}_2^{\text{in}} \\ \vdots \end{bmatrix} = W \begin{bmatrix} \widehat{a}_0 \\ \widehat{a}_1 \\ \vdots \end{bmatrix} \qquad (8)$$

for some unitary matrix $W$. This sequence can be chosen so that $\kappa_k^{\text{in}}$ couple to OQS sequentially, one after the other, at discrete time moments $t_k^{\text{in}}$. Let us denote $m_{\text{in}}(t)$ the number of modes which are already coupled to OQS at the time moment $t$. The function $m_{\text{in}}(t)$ is piecewise constant and increases by jumps at the time moments $t_k^{\text{in}}$ when a new mode is coupled. For convenience of notation we define $m_{\text{in}}(t)$ to be left-continuous: $m_{\text{in}}(t_k^{\text{in}} + 0) = m_{\text{in}}(t_k^{\text{in}}) = m_{\text{in}}(t_k^{\text{in}} - 0) + 1$. Let us discuss the expected rate of such coupling events. The interaction quench at $t = 0$ (i.e. sudden switching-on of the coupling $\widehat{V}^\dagger \widehat{a}_0 + \widehat{V} \widehat{a}_0^\dagger$) is a broadband event. Therefore, we expect that near $t = 0$ the modes $\kappa_k^{\text{in}}$ will appear at a high rate. But fast enough the high-frequency part of the quench spectrum will become decoupled [43, 45], and we reach the Kotelnikov asymptotic rate $t_{k+1}^{\text{in}} - t_k^{\text{in}} \propto \tau = 1/B$, see Fig. 3. We call $\kappa_k^{\text{in}}$ the *incoming Kotelnikov modes*.

If we find these $\kappa_k^{\text{in}}$, then the joint state $|\Psi(t)\rangle$ is found to effectively evolve according to the Schrodinger equation in the rotated frame

$$i\partial_t |\Psi(t)\rangle$$
$$= \left\{ \widehat{H}_s + \sum_{l=1}^{m_{\text{in}}(t)} \left( \widehat{V}^\dagger \chi_l^{\text{in}}(t) \widehat{\kappa}_l^{\text{in}} + \widehat{V} \chi_l^{\text{in}*}(t) \widehat{\kappa}_l^{\text{in}\dagger} \right) \right\} |\Psi(t)\rangle, \qquad (9)$$

where we introduce the creation $\widehat{\kappa}_k^{\text{in}\dagger}$ and annihilation $\widehat{\kappa}_k^{\text{in}}$ operators corresondding to the mode $\kappa_k^{\text{in}}$, and these modes are independent DoFs: $\left[ \widehat{\kappa}_k^{\text{in}}, \widehat{\kappa}_l^{\text{in}\dagger} \right]_\xi = \delta_{kl}$. We assume $\hbar = 1$ in this work. The coupling constant

$$\chi_l^{\text{in}}(t) = \left[ \widehat{a}_0(t), \widehat{\kappa}_l^{\text{in}\dagger} \right]_\xi \qquad (10)$$

is the amplitude to absorb a quantum from the mode $\kappa_l^{\text{in}}$ at time $t$. The time-dependent constant $\chi_l^{\text{in}}(t)$ is a model parameter, which should be found only once for a given $J(\omega)$, see sec. VD below. The Hamiltonian eq. (9) is obtained from eqs. (1)-(2) via the change of frame $W$, eq. (8), in the interaction picture with respect to the free motion of environment.

In Fig. 4 we provide an example calculation for the model

$$\widehat{H}(t) = \varepsilon_s \widehat{\sigma}_+ \widehat{\sigma}_- + \widehat{\sigma}_x f(t) + h \widehat{\sigma}_+ \widehat{a}_0 + h \widehat{\sigma}_- \widehat{a}_0^\dagger$$
$$+ \sum_{i=0}^{\infty} \left\{ \varepsilon \widehat{a}_i^\dagger \widehat{a}_i + h \widehat{a}_{i+1}^\dagger \widehat{a}_i + h \widehat{a}_i^\dagger \widehat{a}_{i+1} \right\}. \qquad (11)$$

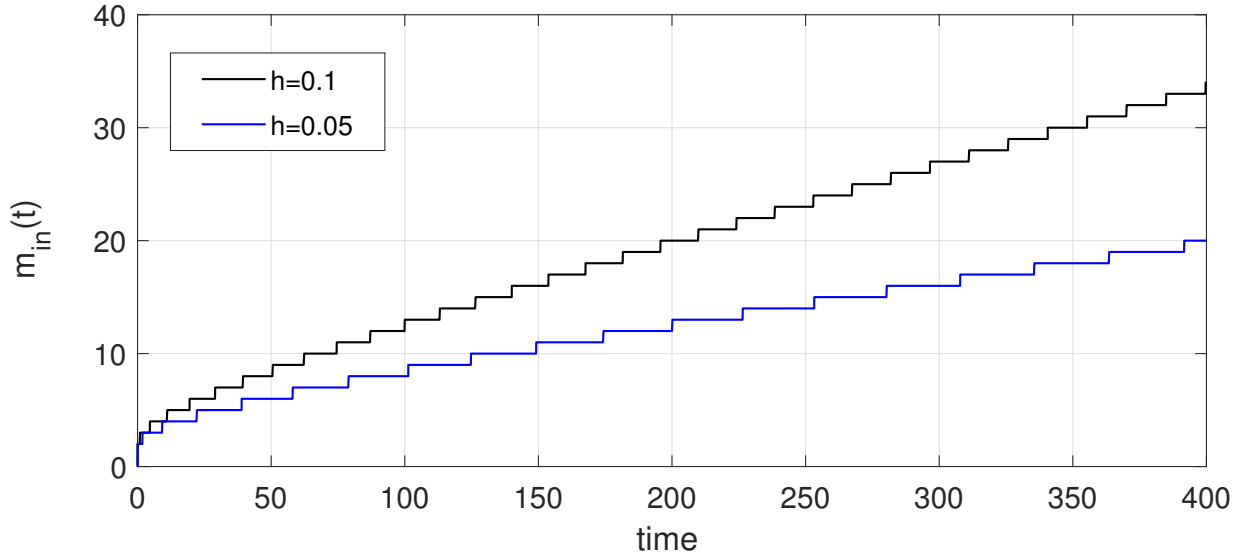

Figure 3. When the quantum noise has a finite bandwidth $B$, its degrees of freedom also appear at discrete time moments $\tau \propto 1/B$. Here we plot the number $m_{\text{in}}(t)$ of coupled incoming quantum Kotelnikov modes $\kappa_k^{\text{in}}$ as a function of time. The case of semicircle spectral density $J(\omega)$ of semi-infinite bosonic chain with hopping $h$ and on-site energy $\varepsilon$, eq. (11). Here the spectral density occupies a band $[\varepsilon - 2h, \varepsilon + 2h]$. Black line: $\varepsilon = 1$, $h = 0.1$. Blue line: $\varepsilon = 1$, $h = 0.05$. New significant modes occurs at discrete time moments. The rate of occurence of new modes is asymptotically constant and proportional to the bandwidth. At small times the rate of occurence is high since the initial coupling quench is a wideband event.

Here the driving classical force $f(t) = 0.1 \cos t$. The chain parameters are $h = 0.05$ and $\varepsilon = 1$. The qubit transition frequency $\varepsilon_s = 1$. The Schrodinger equation in the frame of incoming modes, eq. (9), yields the same results as the numerically exact solution.

The evolution under eq. (9) results in a discrete progressive spread of entanglement which we call the *forward Kotelnikov lightcone*.

Observe that we have a full access to the observables of the environment since the DoFs $\widehat{\kappa}_l^{\text{in}}, \widehat{\kappa}_l^{\text{in}\dagger}$ are true physical DoFs of the environment in the rotated frame eq. (8).

### B. Backward quantum Kotelnikov sampling

Observe that our arrangement $\kappa_1^{\text{in}}, \kappa_2^{\text{in}}, \ldots$ of the DoFs behaves asymmetrically in time. Since the quantum mechanics has the time-reversal symmetry, we expect that the time-reversed process is also happening: at time intervals $\propto 1/B$ some modes $\kappa^{\text{out}}$ get irreversibly decoupled from OQS. We call them the *outgoing Kotelnikov modes*. Here we should distinguish between the two cases: (i) the modes which never become coupled to OQS, and (ii) the modes which couple to OQS but after some time irreversibly decouple. We are not interested in (i): we discard them from our model. However we need to track the evolution of (ii). This consideration determines the way we find $\kappa_l^{\text{out}}$: the mode $\kappa_l^{\text{out}}$ which decouples at time moment $t_l^{\text{out}}$ should be a linear combination of $\kappa_1^{\text{in}} \ldots \kappa_{m(t_l^{\text{out}})}^{\text{in}}$. In other words, the newly decoupled mode should emerge in the subspace of modes that are coupled to OQS by the time $t_l^{\text{out}}$. This is because the modes which are not included in this linear combination have not yet been coupled to OQS.

This picture leads to the following iterative construction of the stream of outgoing modes. The first outgoing mode $\kappa_1^{\text{out}}$ is found by switching to the frame where it appears as an independent DoF,

$$
\begin{bmatrix}
\kappa_1^{\text{out}\dagger} \\
\widehat{\kappa}_1^{\text{rel}\dagger} \\
\vdots \\
\widehat{\kappa}_{m(t_1^{\text{out}})-1}^{\text{rel}\dagger}
\end{bmatrix}
= U_1
\begin{bmatrix}
\widehat{\kappa}_1^{\text{in}\dagger} \\
\vdots \\
\kappa_{m(t_1^{\text{out}})}^{\text{in}\dagger}
\end{bmatrix}
\tag{12}
$$

where $U_1$ is $m_{\text{in}}(t_1^{\text{out}}) \times m_{\text{in}}(t_1^{\text{out}})$ unitary matrix. Here $m_{\text{in}}(t_1^{\text{out}})$ is the number of modes which have coupled before the time of the first decoupling $t_1^{\text{out}}$. We place the new outgoing mode in the first component of the vector on the left hand side. Observe that the remaining $m_{\text{in}}(t_1^{\text{out}}) - 1$ modes can also be rotated somehow. We call them the *relevant modes* since they are coupled to OQS thus are important for the future evolution. We mark them with a superscript "rel". The subsequent outgoing modes are found recurrently. Namely, by the time moment $t_k^{\text{out}}$ we have $k - 1$ outgoing modes $\kappa_1^{\text{out}} \ldots \kappa_{k-1}^{\text{out}}$ and $r(t_k^{\text{out}})$ relevant modes $\widehat{\kappa}_1^{\text{rel}} \ldots \widehat{\kappa}_{r(t_k^{\text{out}})}^{\text{rel}}$, which obvi-

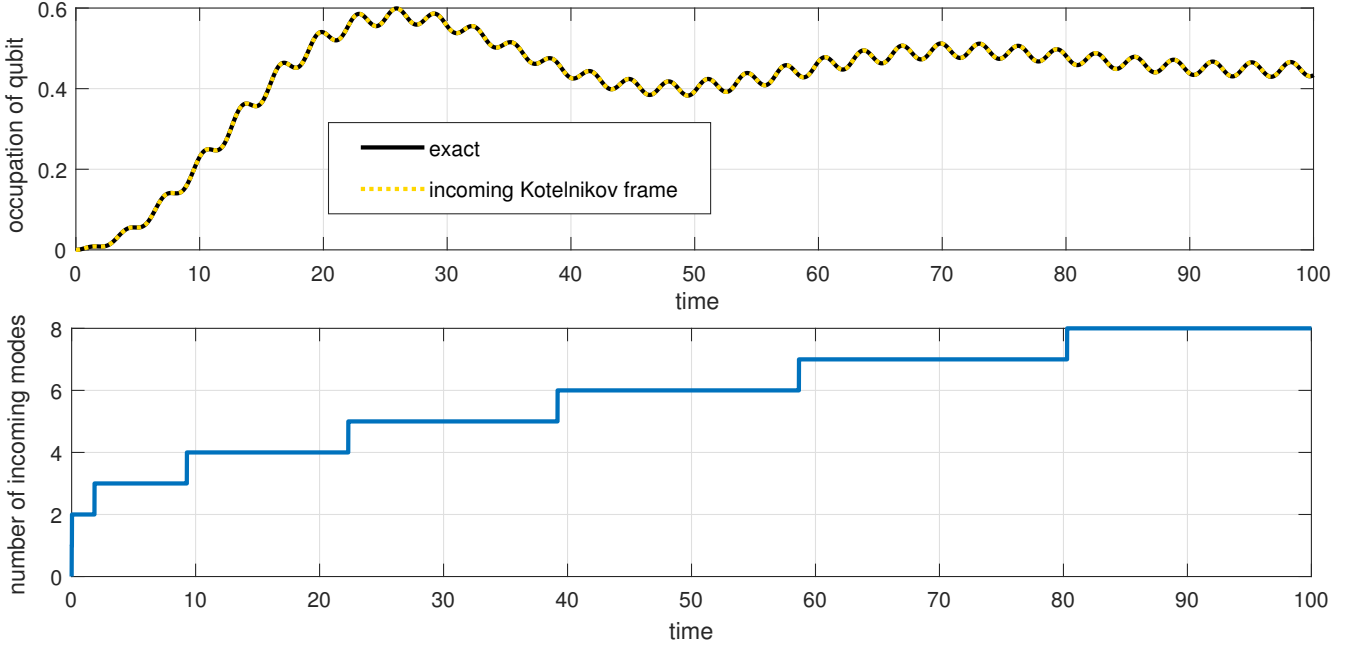

Figure 4. The Schrödinger equation in the frame of incoming Kotelnikov modes eq. (9) yields the same results as a direct numerically exact solution of the Schrödinger equation. Here we consider the model of a driven qubit coupled to a semi-infinite bosonic chain, eq. (11). Upper panel: black curve is the numerically exact solution in a Hilbert space which is truncated in the number of bosonic sites $n = 7$ and in the total occupation of bosonic chain $N = 14$. The parameters $n$ and $N$ where varied until convergence on the presented time interval. Yellow dotted curve is the solution of the same problem in the frame of incoming Kotelnikov modes, eq. (9). Lower panel: the plot of $m_{\mathrm{in}}(t)$ - the number of incoming Kotelnikov modes which are present in the Schrödinger equation at the time moment $t$.

ously satisfy

$$r\left(t_k^{\mathrm{out}}\right) + k - 1 = m_{\mathrm{in}}\left(t_k^{\mathrm{out}}\right). \qquad (13)$$

Then the newly formed outgoing mode is a linear combination of relevant modes:

$$\begin{bmatrix} \widehat{\kappa}_k^{\mathrm{out}\dagger} \\ \widehat{\kappa}_1^{\mathrm{rel}\dagger} \\ \vdots \\ \widehat{\kappa}_{r\left(t_k^{\mathrm{out}}\right)-1}^{\mathrm{rel}\dagger} \end{bmatrix} = U_k \begin{bmatrix} \widehat{\kappa}_1^{\mathrm{rel}\dagger} \\ \vdots \\ \widehat{\kappa}_{r\left(t_k^{\mathrm{out}}\right)}^{\mathrm{rel}\dagger} \end{bmatrix}, \qquad (14)$$

were $U_k$ is $r\left(t_k^{\mathrm{out}}\right) \times r\left(t_k^{\mathrm{out}}\right)$ unitary matrix. We get a picture in which the coupling and decoupling events are interspersed. In the course of a coupling event at $t = t_k^{\mathrm{in}}$, the number of relevant modes increases by one: $r\left(t_k^{\mathrm{out}} + 0\right) = r\left(t_k^{\mathrm{out}}\right) = r\left(t_k^{\mathrm{out}} - 0\right) + 1$, with the new mode $\widehat{\kappa}_{r\left(t_k^{\mathrm{out}}\right)+1}^{\mathrm{rel}} \equiv \widehat{\kappa}_{m\left(t_1^{\mathrm{in}}\right)}^{\mathrm{in}}$. In the course of decoupling event $t = t_k^{\mathrm{out}}$, we rotate the frame, eq. (14).

An important feature of this picture is that all these events have a finite duration $\propto \tau = 1/B$. Therefore, we can do all the rotations smoothly in conjunction with the Hamiltonian evolution. Namely, let us introduce a time interval $[t_k^*, t_k^{\mathrm{out}}]$, where $t_k^*$ is the time of the last event (whatever it is) before $t_k^{\mathrm{out}}$. All the time axis $[0, +\infty)$ becomes covered with such intervals: every time moment

$t$ belongs to some $[t_k^*, t_k^{\mathrm{out}}]$. We can introduce the generator $D(t)$ of $U_k = \exp\left(-i\left(t_k^{\mathrm{out}} - t_k^*\right) D(t)\right)$, where $U_k$ is from eq. (14). Here for convenience of notation we define $D(t)$ as a piecewise constant function on intervals $[t_k^*, t_k^{\mathrm{out}}]$. Then the joint state $|\Psi(t)\rangle$ evolves on $[t_k^*, t_k^{\mathrm{out}}]$ according to the Schrödinger equation

$$i\partial_t |\Psi(t)\rangle = \widehat{H}_{\mathrm{eff}}(t) |\Psi(t)\rangle, \qquad (15)$$

with the effective Hamiltonian in the moving frame

$$\widehat{H}_{\mathrm{eff}}(t) = \widehat{H}_{\mathrm{s}} + \sum_{l=1}^{r(t)} \left( \widehat{V}^\dagger \chi_l^{\mathrm{rel}}(t) \widehat{\kappa}_l^{\mathrm{rel}} + \widehat{V} \chi_l^{\mathrm{rel}*}(t) \widehat{\kappa}_l^{\mathrm{rel}\dagger} \right)$$
$$- \sum_{kl=1}^{r(t)} D_{lk}(t) \widehat{\kappa}_k^{\mathrm{rel}\dagger} \widehat{\kappa}_l^{\mathrm{rel}}. \qquad (16)$$

Here the first line describes the interaction of OQS with $r$ relevant modes, and the coupling constants

$$\chi_l^{\mathrm{rel}}(t) = \left[\widehat{a}_0(t), \widehat{\kappa}_l^{\mathrm{rel}\dagger}\right]_\xi \qquad (17)$$

represent the amplitude to absorb a quantum from the relevant mode $\kappa_l^{\mathrm{rel}}$. The second line of eq. (16) gradually exercises the frame rotation eq. (14).

The equation eq. (16) is ready for applications. The constants of the model, $\chi_l^{\mathrm{rel}}(t)$ and $D_{kl}(t)$, should be found only once for a given $J(\omega)$. The recipe is presented in the following sections.

In Fig. 5 we present example calculation with eqs. (15)-(16) of the same model eq. (11). The exact solution when available coincides with our model. The numbers of coupled modes $m_{\mathrm{in}}(t)$, emerged outgoing modes $m_{\mathrm{out}}(t)$, and the relevant modes $r(t)$ are presented on Fig. 6.

The appearance of the irreversibly decoupled modes results in a discrete progressive *disentanglement* structure which we call the *backward Kotelnikov lightcone*.

### C. Tracing out the irreversibly decoupled modes

The merit of the presented model is that the irreversibly decoupled modes can be traced out as soon as they appear during the real time motion:

$$\widehat{\rho}_{\mathrm{rel}}(t) = \mathrm{Tr}_{\kappa_1^{\mathrm{out}}\ldots\kappa_k^{\mathrm{out}}}\left\{|\Psi(t)\rangle\langle\Psi(t)|\right\}, \quad (18)$$

for all $k: t_k^{\mathrm{out}} < t$. Here the subscript "rel" denotes the fact that the density matrix $\widehat{\rho}_{\mathrm{rel}}(t)$ depends only on the quantum numbers of OQS and of $r(t)$ relevant modes. This is a reduced description of real-time motion of bounded complexity with respect to the time, with a closed equation of motion

$$\partial_t \widehat{\rho}_{\mathrm{rel}}(t) = -i\left[\widehat{\rho}_{\mathrm{rel}}(t), \widehat{H}_{\mathrm{eff}}(t)\right]. \quad (19)$$

### D. Number of relevant modes saturates

In our model the decoupling process is a time-reversed coupling process, and vice versa, see section V C 3. Therefore, we expect them to have the same asymptotic rate $\propto B$: the number $m_{\mathrm{in}}(t)$ of coupled and the number $m_{\mathrm{out}}(t)$ of decoupled modes grow with the same rate. As a result, their difference, the number of relevant modes $r(t)$ (see eq. (13)), is asymptotically constant: $r(t) \propto r$, see Fig. 6. Therefore, the von Neumann equation (19) provides a representation of real-time motion with bounded number of DoFs on wide time scales.

### E. Monte Carlo sampling of quantum jumps

There is alternative to the partial trace in eq. (18). When a new outgoing mode $\kappa_k^{\mathrm{out}}$ is formed at $t = t_k^{\mathrm{out}}$, we can apply the von Neumann measurement model [64]: the mode $\kappa_k^{\mathrm{out}}$ is an auxiliary probe which was previously coupled to OQS. It is entangled with OQS according to

the Schmidt decomposition

$$\begin{aligned}&|\Psi\left(t_k^{\mathrm{out}}\right)\rangle \\ &= \sum_p c_p(k)\left|\Psi_{\mathrm{collapsed}}^{(p)}\left(t_k^{\mathrm{out}}\right)\right\rangle_{\mathrm{rel}} \otimes \left|\mathrm{jump}^{(p)}\left(t_k^{\mathrm{out}}\right)\right\rangle_{\kappa_k^{\mathrm{out}}},\end{aligned} \quad (20)$$

were we consider a biparticle system: one part labeled "rel" is the open system with relevant modes, and the second part labeled $\kappa_k^{\mathrm{out}}$ is the newly formed outgoing mode. Since the probe $\kappa_k^{\mathrm{out}}$ is irreversibly decoupled, the entanglement structure (20) is invariant under the future evolution. Then, according to the von Neumann measurement model, we can interpret eq. (20) as a $p$th quantum jump at time $t = t_k^{\mathrm{out}}$,

$$|\Psi\left(t_k^{\mathrm{out}}\right)\rangle \to \left|\Psi_{\mathrm{collapsed}}^{(p)}\left(t_k^{\mathrm{out}}\right)\right\rangle_{\mathrm{rel}}, \quad (21)$$

with a probability $P_k(p) = |c_p(k)|^2$. These jumps are irreversible and non-Markovian since they happen during a finite time interval $t_k^{\mathrm{out}} - t_k^* \propto 1/B$.

At a time $t$ there are $m_{\mathrm{out}}(t)$ decoupled modes. Each of them is accompanied with a quantum jump, which are obtained by a recurrent application of the von Neumann rule eq. (20)-(21). Therefore, before the time $t$ there are $m_{\mathrm{out}}(t)$ jumps. They are characterized by a history $h$ of choices, $h = (p_1, p_2, \ldots, p_k) = (p_k)_{k: t_k^{\mathrm{out}} \leq t}$, which occurs with the probability $P(p_1 p_2 \ldots p_k) = \prod_{k: t_k^{\mathrm{out}} \leq t}|c_{p_k}(k)|^2$. Then the reduced density matrix eq. (18) is computed as an average over the histories $h$ of jumps before $t$,

$$\begin{aligned}&\widehat{\rho}_{\mathrm{rel}}(t) = \\ &\overline{\left[\left|\Psi_{\mathrm{collapsed}}^{(h)}(t)\right\rangle_{\mathrm{rel}}\,\,_{\mathrm{rel}}\left\langle\Psi_{\mathrm{collapsed}}^{(h)}(t)\right|\right]}^{\mathrm{over\ all}\ h}. \quad (22)\end{aligned}$$

Here by $\left|\Psi_{\mathrm{collapsed}}^{(h)}(t)\right\rangle_{\mathrm{rel}}$ we denote a particular trajectory of jumps,

$$\begin{aligned}|\Psi\left(t_1^{\mathrm{out}}\right)\rangle &\to \left|\Psi_{\mathrm{collapsed}}^{(p_1)}\left(t_1^{\mathrm{out}}\right)\right\rangle_{\mathrm{rel}}, \\ \left|\Psi_{\mathrm{collapsed}}^{(p_1)}\left(t_2^{\mathrm{out}}\right)\right\rangle_{\mathrm{rel}} &\to \left|\Psi_{\mathrm{collapsed}}^{(p_1 p_2)}\left(t_2^{\mathrm{out}}\right)\right\rangle_{\mathrm{rel}}, \\ \left|\Psi_{\mathrm{collapsed}}^{(p_1 p_2)}\left(t_3^{\mathrm{out}}\right)\right\rangle_{\mathrm{rel}} &\to \left|\Psi_{\mathrm{collapsed}}^{(p_1 p_2 p_3)}\left(t_3^{\mathrm{out}}\right)\right\rangle_{\mathrm{rel}}, \\ &\mathrm{etc.} \quad (23)\end{aligned}$$

This procedure results in a novel Monte Carlo simulation technique of real-time motion. It has a benefit with respect to the density matrix formulation (18)-(19) since the density matrix has a quadratically higher dimension than that of wavefunction, $\dim \widehat{\rho}_{\mathrm{rel}}(t) = \left(\dim\left|\Psi_{\mathrm{collapsed}}^{(h)}(t)\right\rangle_{\mathrm{rel}}\right)^2$.

Actually the Fig. 5 was computed with this technique. The averaging was done over 36000 randomly generated jump histories. In Fig. 7 we present an example of single random history of jumps. It is seen that the qubit observable behaves discontinuously at the times of jumps (when $m_{out}(t)$ increases by 1).

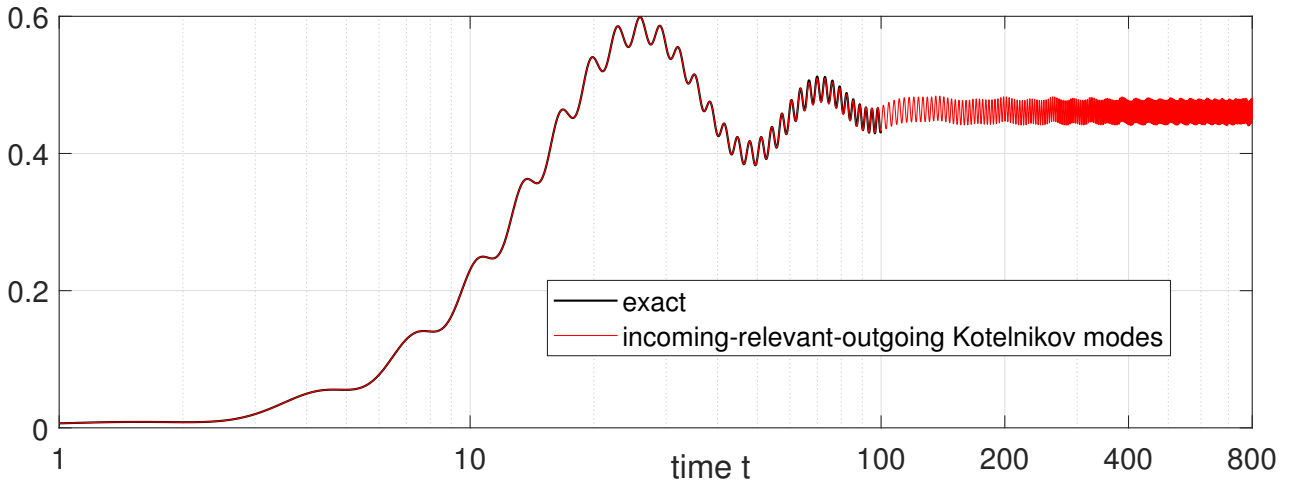

Figure 5. The plot of the qubit occupation for the model eq. (11). The Schrodinger equation in the frame of incoming/relevant/outgoing Kotelnikov modes eq. (15)-(16) yields the same results as a direct numerically exact solution of the Schrodinger equation. We consider the same case as in Fig. 4. The outgoing modes were traced out via the quantum jump Monte Carlo simulation from section IV E. The Fock space of relevant modes was truncated at a maximal occupation of 6 quanta.

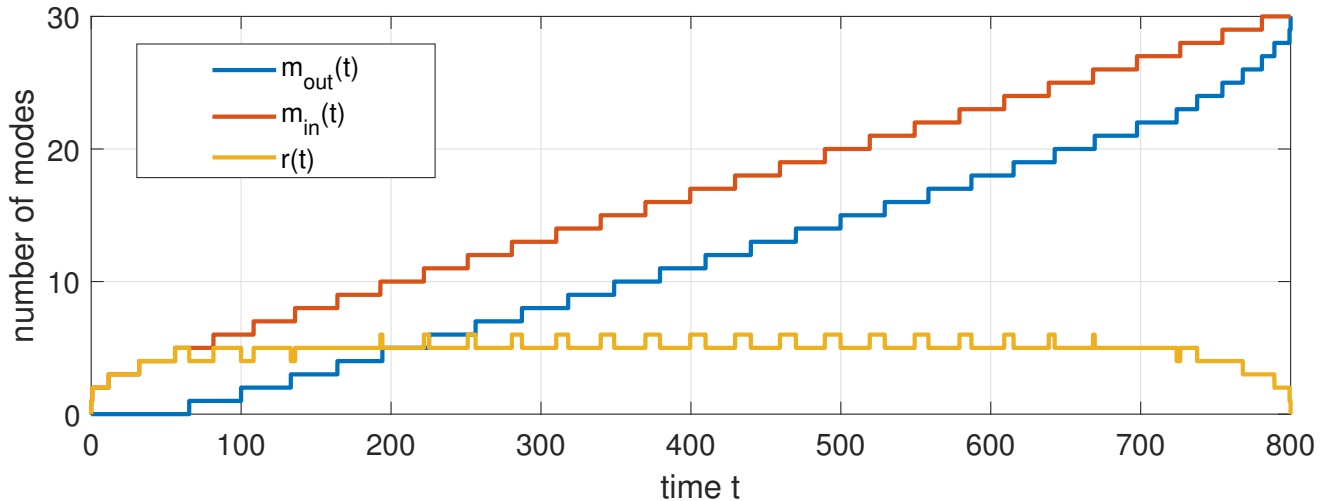

Figure 6. The number of coupled modes $m_{in}(t)$ and the number of outgoing modes $m_{out}(t)$ grow in time with the same asymptotic rate $\propto B$, where $B$ is the bandwidth of the spectral density $J(\omega)$. Therefore, the number of relevant modes $r(t) = m_{in}(t) - m_{out}(t)$ is asymptotically constant in time. The case of semicircle spectral density $J(\omega)$ of semiinfinite bosonic chain with hopping $h$ and on-site energy $\varepsilon$, eq. (11). Here the spectral density occupies a band $[\varepsilon - 2h, \varepsilon + 2h]$. Here $\varepsilon = 1$, $h = 0.05$. The relative significance treshold $r_{cut} = 10^{-4}$.

This scheme is also of conceptual interest since it represents the environment as a measurement apparatus which autonomously selects the time moment of measurement and the preferred basis, without the intervention of the human experimenter. The classical reality (a statistical ensemble of jump histories) is encoded in the emerging invariant entanglement structure.

### F. Bounded population of relevant modes

The outgoing modes $\kappa_k^{out}$ carry away quantum excitations. Therefore, we have two fluxes of quanta: the incoming flux $j^{in}$ of quanta which are emitted by OQS into the relevant $\kappa_l^{rel}$ modes, and the outgoing flux $j^{out}$ quanta which are irreversibly carried away by $\kappa_k^{\text{out}}$. On a physical grounds it is natural to conjecture a balance between these two fluxes. Our calculation supports this conjecture, Fig. 8.

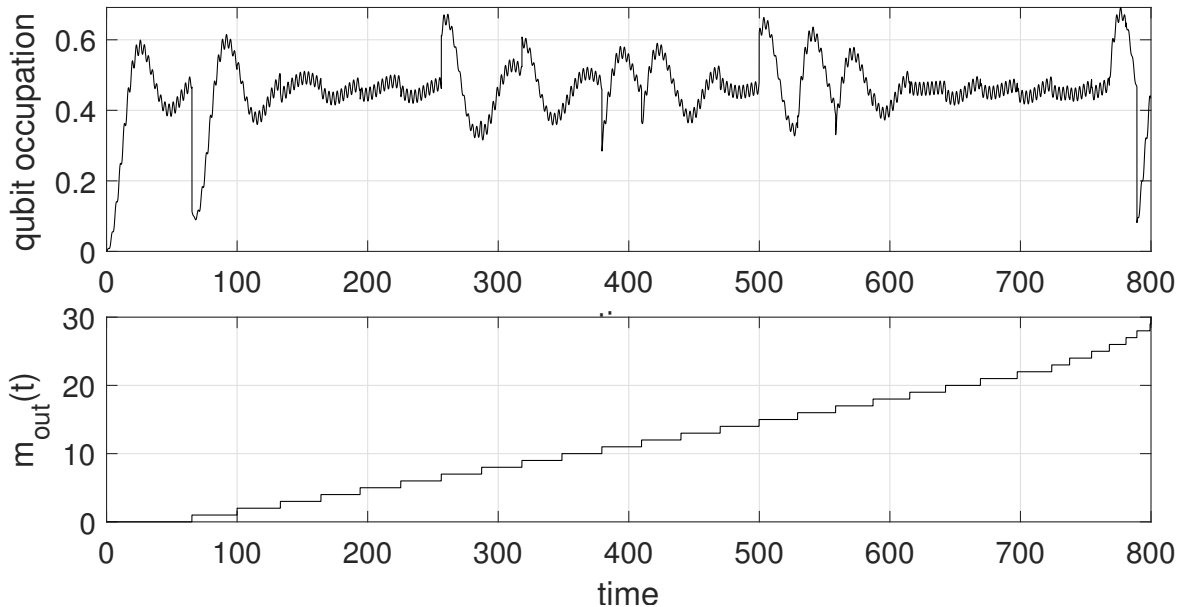

Figure 7. An example of a single jump history: the qubit observable (OQS) experiences discontinuities due to quantum jumps. The latter happen at time moments $t_k^{\text{out}}$ when $m_{\text{out}}(t)$ increases by 1. The model is eq. (11). The Fock space of relevant modes was truncated at a maximal occupation of 6 quanta.

### G. Geometric arrow of time

Suppose we know that the environment is in some global equilibrium state $\widehat{\rho}_{\text{b}}^{\text{eq}}$, and that OQS was coupled to the environment at some time moment in the past. Then our streams of incoming and outgoing Kotelnikov modes define an arrow of time. Indeed, at any time moment $t$ the future incoming modes $\kappa_k^{\text{in}}$, $t_k^{\text{in}} > t$, did not have an opportunity to significantly interact with OQS. Therefore, the partial density matrix of the incoming stream should coincide with the equilibrium one at all times:

$$\widehat{\rho}_{\text{in}}(t) \equiv \text{Tr}_{\text{all } \kappa_k^{\text{in}}:\, t_k^{\text{in}} \leq t} \text{Tr}_{\text{S}} \{\widehat{\rho}(t)\}$$
$$= \text{Tr}_{\text{all } \kappa_k^{\text{in}}:\, t_k^{\text{in}} \leq t} \{\widehat{\rho}_{\text{b}}^{\text{eq}}\} \equiv \widehat{\rho}_{\text{in}}^{\text{eq}}(t), \quad (24)$$

where $\widehat{\rho}(t)$ is the joint density of OQS and environment due to the coupling quench. At the same time, the outgoing stream carries away the disturbance due to the coupling quench, therefore its partical density matrix in general differs from the equilibrium one:

$$\widehat{\rho}_{\text{out}}(t) \equiv \text{Tr}_{\text{all} \kappa_k^{\text{out}}:\, t_k^{\text{out}} \geq t} \text{Tr}_{\text{S}} \{\widehat{\rho}(t)\}$$
$$\neq \text{Tr}_{\text{all} \kappa_k^{\text{out}}:\, t_k^{\text{out}} \geq t} \{\widehat{\rho}_{\text{b}}^{\text{eq}}\} \equiv \widehat{\rho}_{\text{out}}^{\text{eq}}(t). \quad (25)$$

We show in section V C 3 that under the time reversal $\widehat{\rho}_{\text{in}}(t)$ and $\widehat{\rho}_{\text{out}}(t)$ are swapped. However, in the latter case $\widehat{\rho}_{\text{in}}(t) \neq \widehat{\rho}_{\text{in}}^{\text{eq}}$, so that we detect the "wrong" time direction.

We call this arrow of time geometric because it is revealed by a specific arrangement of the microscopic degrees of freedom. By itself, this arrangement is time-symmetric ($\kappa_k^{\text{out}}$ and $\kappa_k^{\text{in}}$ swap under the time reversal). However the boundary condition (global equilibrium of uncoupled environment at $t = -\infty$) is asymmetric, which leads to the emergence of the preferred time direction.

### H. Entropy of the ensemble of classical records

As we have discussed above, the frame of outgoing modes $\kappa_k^{\text{out}}$ reveals the emerging asymptotically invariant entanglement structure:

$$|\Psi(t)\rangle = \sum_{p_1 \ldots p_k} c_{p_1}(1) \ldots c_{p_k}(k) \left| \Psi_{\text{collapsed}}^{(p_1 \ldots p_k)}(t) \right\rangle_{\text{rel}}$$

$$\otimes \left| \text{jump}^{(p_1)}\left(t_1^{\text{out}}\right) \right\rangle_{\kappa_1^{\text{out}}} \ldots \otimes \left| \text{jump}^{(p_k)}\left(t_k^{\text{out}}\right) \right\rangle_{\kappa_k^{\text{out}}} \quad (26)$$

for $t_k^{\text{out}} \leq t \leq t_{k+1}^{\text{out}}$. This structure encodes the classical ensemble of jump histories $(p_1 \ldots p_k)$ with probabilities

$$P(p_1 p_2 \ldots p_k) = \prod_{k:\, t_k^{\text{out}} \leq t} |c_{p_k}(k)|^2. \quad (27)$$

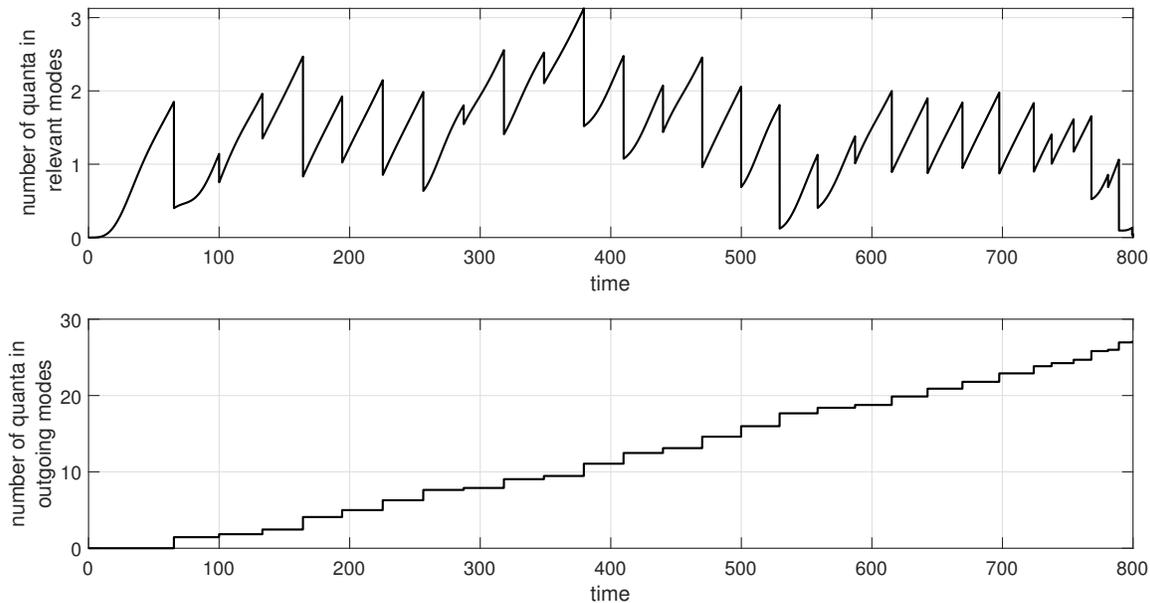

Figure 8. Here we present a single random history of quantum jumps. Upper plot: total occupation of relevant modes $\kappa_k^{\mathrm{rel}}$. Lower plot: total occupation of outgoing modes $\kappa_k^{\mathrm{out}}$. The model is eq. (11). We see that the occupation of relevant modes is bounded. Since the number of relevant modes is bounded, and their population is bounded, then the Fock space of relevant modes can be truncated by keeping only a finite number $d$ of basis states. As a result, the quantum evolution becomes a matrix product state whose bond dimension $d$ is bounded with time.

The entropy of such ensemble is

$$S_{\mathrm{jump}}\left(\Psi\left(t\right)\right) = -\sum_{p_1 \ldots p_k} P\left(p_1 \ldots p_k\right) \ln P\left(p_1 \ldots p_k\right).$$

(28)

Informally, we can say that for a given pure state $|\Psi\left(t\right)\rangle$, the entropy $S_{\mathrm{jump}}\left(\Psi\left(t\right)\right)$ quantifies how much classical reality was generated in this pure state due to non-Markovian decoherence.

## V. VARIATIONAL PRINCIPLE AND NUMERICAL SCHEME

In this section we provide a recipe how to compute the constants $\chi_l^{\mathrm{rel}}\left(t\right)$ and $D_{lk}\left(t\right)$ of the model eq. (15)-(16). First we introduce a variational principle for the real time motion which yields the computational scheme which in its turn is presented at the end of this section.

We begin by noting that the classical Kotelnikov theorem eq. (5) can be reformulated in a statistical form: on every finite time interval $T$ there is a finite number $m\left(t\right) \propto BT$ of independent statistically significant DoFs for a signal with effective bandwidth $B$. See appendix A. Also, this statistical reformulation yields the desired causal variant of the Kotelnikov theorem.

Thererore, we proceed by constructing measures of statistical significance of environment DoFs on time intervals.

### A. Measure of retarded statistical significance

Returning to the coupling quench of OQS, we face the problem how to identify the statistically significant DoFs of environment on a time interval $[0, t]$. We need to identify these DoFs *before* sovling the many body problem eq.(1). The simplest trick we can do is to replace OQS in eq. (1) with a classical white noise $\xi\left(t\right)$:

$$\widehat{H} \rightarrow \widehat{H}_{\xi+}\left(t\right) = \xi\left(t\right)^* \widehat{a}_0\left(t\right) + \xi\left(t\right) \widehat{a}_0^\dagger\left(t\right)$$

(29)

(here we consider a specific case of bosonic environment), $\overline{\xi^*\left(t\right)\xi\left(t'\right)} = \delta\left(t - t'\right)$. Since the white noise $\xi\left(t\right)$ has infinite bandwidth and uniform spectral density, it efficiently lights up all the available DoFs of the environment. The intensity of such interaction is modulated with $J\left(\omega\right)$. Then, given a trial environment DoF $\phi$, with the corresponding creation operator $\widehat{\phi}^\dagger = \sum_{i=1}^{\infty} \langle i | \phi \rangle \widehat{a}_i^\dagger$, we can estimate its statistical significance in the following way. We compute its average occupation $n_+\left(t; \phi\right)$ by averaging over the ensemble of noisy evolutions of the environment $|\Psi_{\xi+}\left(t\right)\rangle_{\mathrm{b}}$ under the Hamiltonian eq. (29),

$$n_+\left(t; \phi\right) = \langle \phi | \widehat{\rho}_+\left(t\right) | \phi \rangle,$$

(30)

where the one-body reduced density matrix $\widehat{\rho}_+\left(t\right)$ is $\langle i | \widehat{\rho}_+\left(t\right) | j \rangle = \overline{\langle \Psi_{\xi+}\left(t\right) | \widehat{a}_i^\dagger \widehat{a}_j | \Psi_{\xi+}\left(t\right)\rangle}$. Here for each noise realization we solve the Schrodinger equation $i\partial_t |\Psi_{\xi+}\left(t\right)\rangle_{\mathrm{b}} = \widehat{H}_{\xi+}\left(t\right) |\Psi_{\xi+}\left(t\right)\rangle_{\mathrm{b}}$ with the vacuum initial condition $|\Psi_{\xi+}\left(0\right)\rangle_{\mathrm{b}} = |0\rangle_{\mathrm{b}}$. If $n_+\left(t; \phi\right)$ is negligible, then no other open system can excite this DoF, and

$\phi$ can be safely discarded. Indeed, unlike white noise, OQS interacts only with those DoFs which are close to its transition frequencies. Therefore, OQS is a less efficient illuminator than the white noise.

We call $n_+ (t; \phi)$ the measure of retarded statistical significance, i.e. the statistical significance of $\phi$ for the past quantum motion up to time $t$. The modes $\phi$ with negligible $n_+ (t; \phi)$ can be discarded from the Hamiltonian $\widehat{H}_{\mathrm{b}}$ when solving the Schrödinger equation on $[0, t]$.

In appendix B we show that $\widehat{\rho}_+ (t)$ has a simple closed solution

$$\langle i | \widehat{\rho}_+ (t) | j \rangle = \int_0^t d\tau \phi_i^* (\tau) \, \phi_j (\tau) , \qquad (31)$$

where $\phi_j (\tau)$ is a free one-particle wavepacket in the environment, which is initially located on the site 0 of the environment (which is coupled to OQS), $\phi_j (0) = \delta_{j0}$, and at later times its free motion under $\widehat{H}_{\mathrm{b}}$ is given by a first-quantized Schrödinger equation

$$\partial_\tau \phi_j (\tau) = i\varepsilon_j \phi_j (\tau) + i h_j \phi_{j+1} (\tau) + i h_{j-1} \phi_{j-1} (\tau) . \qquad (32)$$

These equations can be easily solved on a computer.

For a given arbitrary trial mode $\phi$, the measure $n_+ (t; \phi)$ is a monotonically increasing function of $t$, which follows from eq. (31). Therefore, given a significance treshold $r_{\mathrm{cut}}$, there is a unique time moment $t_\phi^{\mathrm{in}} \in [0, \infty]$ such that before $t_\phi^{\mathrm{in}}$ the mode $\phi$ is effectively decoupled,

$$n_+ (t; \phi) < r_{\mathrm{cut}} \text{ for } t < t_\phi^{\mathrm{in}}, \qquad (33)$$

and becomes coupled after this time moment,

$$n_+ (t; \phi) > r_{\mathrm{cut}} \text{ for } t > t_\phi^{\mathrm{in}}. \qquad (34)$$

If the mode $\phi$ is always significant, we formally put $t_\phi^{\mathrm{in}} = 0$. If the mode $\phi$ is never significant, we formally put $t_\phi^{\mathrm{in}} = +\infty$.

### B. Measure of advanced statistical significance

Let us assume that we consider the quantum motion on a maximal time interval $[0, T]$. Analogously to the previous section, we can introduce the measure of statistical significance for the future quantum motion $n_- (t; \phi)$ on the time interval $[t, T]$. We call $n_- (t; \phi)$ the measure of advanced statistical significance.

In order to define $n_- (t; \phi)$, we again employ the white noise trick. We consider the ensemble of noisy evolutions of the environment $|\Psi_{\xi-} (\tau)\rangle_{\mathrm{b}}$, but this time the noise is switched on only after $\tau = t$ :

$$\widehat{H}_{\xi-} (\tau) = \theta (\tau - t) \left\{ \xi (\tau)^* \, \widehat{a}_0 (t) + \xi (\tau) \, \widehat{a}_0^\dagger (t) \right\} . \qquad (35)$$

We solve the Schrödinger equation $i\partial_\tau |\Psi_{\xi-} (\tau)\rangle_{\mathrm{b}} = \widehat{H}_{\xi-} (\tau) |\Psi_{\xi-} (\tau)\rangle_{\mathrm{b}}$ with the vacuum initial condition $|\Psi_{\xi-} (0)\rangle_{\mathrm{b}} = |0\rangle_{\mathrm{b}}$. Then, given a trial DoF $\phi$, with the corresponding creation operator $\widehat{\phi}^\dagger = \sum_{i=1}^\infty \langle i | \phi \rangle \, \widehat{a}_i^\dagger$, we can estimate its advanced statistical significance as average occupation

$$n_- (t; \phi) = \langle \phi | \widehat{\rho}_- (t) | \phi \rangle , \qquad (36)$$

where the one-body reduced density matrix $\widehat{\rho}_- (t)$ is $\langle i | \widehat{\rho}_- (t) | j \rangle = \langle \Psi_{\xi-} (t) | \widehat{a}_i^\dagger \widehat{a}_j | \Psi_{\xi-} (t) \rangle$. The modes $\phi$ with negligible $n_- (t; \phi)$ can be discarded from the Hamiltonian $\widehat{H}_{\mathrm{b}}$ when solving the Schrödinger equation *after* the time $t$.

Analogously to appendix B, one can show that $\widehat{\rho}_- (t)$ has a simple closed solution

$$\langle i | \widehat{\rho}_- (t) | j \rangle = \int_t^T d\tau \phi_i^* (\tau) \, \phi_j (\tau) , \qquad (37)$$

where $\phi_j (\tau)$ is the same as in eq. (32). This equation can also be easily solved on a computer.

The measure $n_- (t; \phi)$ is a monotonically *decreasing* function of $t$, which follows from eq. (37). Therefore, given a significance treshold $r_{\mathrm{cut}}$, there is a unique time moment $t_\phi^{\mathrm{out}} \in [0, \infty]$ such that *after* $t_\phi^{\mathrm{out}}$ the mode $\phi$ is irreversibly decoupled,

$$n_- (t; \phi) < r_{\mathrm{cut}} \text{ for } t > t_\phi^{\mathrm{out}}, \qquad (38)$$

and is coupled before this time moment,

$$n_- (t; \phi) > r_{\mathrm{cut}} \text{ for } t < t_\phi^{\mathrm{out}}. \qquad (39)$$

If the mode is always significant, we formally put $t_\phi^{\mathrm{out}} = +\infty$. If the mode is never significant, we formally put $t_\phi^{\mathrm{out}} = 0$.

### C. Streams of incoming and outgoing Kotelnikov modes

Having defined meaningful, easily computable measures of significance, the problem of finding the streams of incoming $\kappa_l^{\mathrm{in}}$ and outgoing $\kappa_l^{\mathrm{out}}$ Kotelnikov modes becomes purely technical.

#### 1. Incoming Kotelnikov modes

Given a maximal time interval $[0, T]$, we define the stream of incoming $\kappa_l^{\mathrm{in}}$ modes, $l = 1 \ldots m_{\mathrm{in}} (T)$, as a sequence of independent DoFs such that (I.1) it is the set of all independent DoFs which are significant at $t = T$. That is, for any other trial independent DoF $\phi$ we will necessarily have $n_+ (T; \phi) < r_{\mathrm{cut}}$. (I.2) It is the slowest coupling sequence. That is, we can always rotate the frame via some unitary matrix $U$ and arrange the set $\kappa_l^{\text{in}}$ as another sequence $\kappa_l^{\text{in}\prime} = \sum_k U_{lk}\kappa_k^{\text{in}}$. Then the slowest coupling means that the sequence $\kappa_l^{\text{in}}$ couples at a sequence of times $t_k^{\text{in}}$ which are the most delayed: $t_k^{\text{in}} \geq t_k^{\text{in}\prime}$, where $t_k^{\text{in}\prime}$ are the coupling time for any other arrangement $\kappa_l^{\text{in}\prime}$. We want to have (I.1) in order to have a faithful representation of the quantum motion. We want to have (I.2) because we want to slow down the growth of complexity as much as possible. It is evident that the lowest bound for the growth rate is nothing but the Kotelnikov sampling rate $\propto B$.

### 2. Outgoing Kotelnikov modes

Given a maximal time interval $[0,T]$, we define the stream of outgoing $\kappa_l^{\text{out}}$ modes, $l = 1 \ldots m_{\text{in}}(T)$, as a sequence of independent DoFs such that (O.1) it is the set of all independent DoFs which are significant at $t = T$. That is, for any other trial independent DoF $\phi$ we will necessarily have $n_+(T;\phi) < r_{\text{cut}}$. (O.2) It is the fastest decoupling sequence. That is, we can always rotate the frame via some unitary matrix $U$ and arrange the set $\kappa_l^{\text{out}}$ as another sequence $\kappa_l^{\text{out}\prime} = \sum_k U_{lk}\kappa_k^{\text{out}}$. Then the fastest decoupling means that the sequence $\kappa_l^{\text{out}}$ decouples at a sequence of times $t_k^{\text{out}}$ which are the earliest times: $t_k^{\text{out}} \leq t_k^{\text{out}\prime}$, where $t_k^{\text{out}\prime}$ are the coupling time for any other arrangement $\kappa_l^{\text{out}\prime}$. Again, we want to have (O.1) in order to have a faithful representation of the quantum motion. We want to have (O.2) because we want to speed up the decay of complexity as much as possible. It is evident that the maximal decay rate is bounded from above by bandwidth, i.e. it is the Kotelnikov sampling rate $\propto B$.

### 3. Time reversal symmetry between the streams of incoming and outgoing modes

These definitions are manifestly time reversal invariant. Let is define the time reversal operator $\mathcal{T}$ on the time interval $[0,T]$: $\mathcal{T}[\phi(t)] = \phi^*(T-t)$. Then $\mathcal{T}\widehat{\rho}_\pm(t)\mathcal{T} = \widehat{\rho}_\mp(T-t)$, and the measures of statistical significance swap: $n_\pm(t;\phi) = n_\mp(T-t;\mathcal{T}\phi)$. The slowest coupling becomes the fastest decoupling, and vice versa, so that the incoming and the outgoing Kotelnikov modes also swap: $\mathcal{T}\kappa_l^{\text{in}} = \kappa_{m_{\text{in}}(T)-l+1}^{\text{out}}$, with $t_l^{\text{in}} = T - t_{m_{\text{in}}(T)-l+1}^{\text{out}}$. Observe that the initial condition for eq. (32) is not time reversal invariant. But this does not matter: the time reversal results in a unitarily displaced initial condition. That is, the frame of the whole environment is just unitarily displaced, but the model constants $\chi_l^{\text{rel}}(t)$ and $D_{lk}(t)$ in eq. (16) are invariant under such a change, and $\widehat{H}_{\text{eff}}(t)$ is the same.

### 4. Composition of incoming and outgoing streams

As we pointed out in sec. IV B, we are only interested in the outgoing modes which were previously coupled to OQS. Therefore, we impose an additional constraint on the definition of outgoing modes: (O.3) $\kappa_k^{\text{out}}$ should be a linear combination of all $\kappa_l^{\text{in}}$ with $t_l^{\text{in}} \leq t_k^{\text{out}}$. This definition is not required for the correctness of the model. But it is required to achieve the balance of complexity: due to (O.3) the number of relevant modes saturates, Fig. 6. Observe that formally this requirement breaks the strict time reversal symmetry between the incoming and outgoing streams: $\mathcal{T}\kappa_l^{\text{in}} \neq \kappa_{m_{\text{in}}(T)-l+1}^{\text{out}}$. However this symmetry holds in a more general sense, $\mathcal{T}\kappa_l^{\text{in}} = \kappa_{m_{\text{in}}(T)-l+1}^{\text{out}\prime}$ and $\mathcal{T}\kappa_l^{\text{out}} = \kappa_{m_{\text{in}}(T)-l+1}^{\text{in}\prime}$, where $\kappa_k^{\text{in}\prime}$ and $\kappa_k^{\text{out}\prime}$ are some other incoming and outgoing streams which satisfy the requirements (I.1, O.1) and suboptimally satisfy theirs corresponding requirements (I.2, O.2).

### D. Numerical scheme

#### 1. The preparatory stage.

Given $J(\omega)$ we construct the equivalent semiinfinite chain representation eq. (2). Then the maximal time interval $[0,T]$ is chosen. The one-particle Schrodinger equation eq. (32) is solved for $\phi_j(\tau)$. Due to the Lieb-Robinson bounds the Schrodinger equation can be truncated at a finite number of sites. Since this is a single particle problem, we can keep a huge number of sites and this does not place practical constraints in numerical computations. The density matrix $\widehat{\rho}_+(T)$ is computed using eq. (31). The relative significance treshold is chosen e.g. $r_{\text{cut}} = 10^{-4}$. Then the most significant eigenvalues of $\widehat{\rho}_+(T)$ are found:

$$\widehat{\rho}_+(T)|\kappa_l\rangle = \pi_l|\kappa_l\rangle,\qquad(40)$$

where $|\kappa_l\rangle$ are sorted in the descending order of $\pi_l$ and we keep a finite number $m_{\text{in}}(T)$ of modes which are beyond the significance level, $\pi_k/\pi_1 > r_{\text{cut}}$ for $k = 1 \ldots m_{\text{in}}(T)$. This way we satisfy the requirements (I.1, O.1) for the incoming and outoing modes. The Fig. 3 presents an example of such a calculation.

Usually the number $m_{\text{in}}(T)$ is much smaller then the number of spatial sites we need to keep due to the Lieb-Robinson bounds. Therefore, here and below we assume that we perform the computations in the frame of modes $|\kappa_1\rangle \ldots |\kappa_{m_{\text{in}}(T)}\rangle$. Then $\widehat{\rho}_+(T)$ is a $m_{\text{in}}(T) \times m_{\text{in}}(T)$ matrix.

#### 2. The stream of incoming modes

As we have discussed in sec. V C 3, the slowest coupling is a time-reversed fastest decoupling. Therefore,

we propagate $\widehat{\rho}_+(T)$ backwards in time, from $t = T$ to $t = 0$. Each moment of time we compute its eigenvalues,

$$\widehat{\rho}_+(t)\left|\kappa_l^+(t)\right\rangle = \pi_l^+(t)\left|\kappa_l^+(t)\right\rangle, \qquad (41)$$

where $\left|\kappa_l^+\right\rangle$ are sorted in the descending order of $\pi_l^+$. We remember the times $t_k^{\rm in}$ when the lowest eigenvalue $\pi_k^+(t_k^{\rm in})$ falls below the significance treshold, $\pi_k^+(t_k^{\rm in})/\pi_1^+(t_1^{\rm in}) < r_{\rm cut}$. Each time the corresponding mode $\kappa_k^+(t_k^{\rm in})$ is remembered as the incoming mode, $\kappa_k^{\rm in} \equiv \kappa_k^+(t_k^{\rm in})$. This way we satisfy (I.2). Immediately after that the mode $\kappa_k^{\rm in}$ is excluded from the backpropagation: the dimensions of $\widehat{\rho}_+(t)$ decrease by 1 at each time moment $t_k^{\rm in}$. Recurrently repeating this step we find the incoming modes satisfying (I.1) and (I.2)

### 3. The stream of outgoing modes

This time we propagate the density matrix $\widehat{\rho}_-(t)$ eq. (37) forward in time, from $t = 0$ to $t = T$. Initially $\widehat{\rho}_-(t)$ is a $m_{\rm in}(t) \times m_{\rm in}(t)$ matrix since we project it to the subspace of incoming modes $\kappa_1^{\rm in} \dots \kappa_{m_{\rm in}(t)}^{\rm in}$ which have coupled before the time moment $t$. This way we satisfy (O.3). Each moment of time we compute its eigenvalues,

$$\widehat{\rho}_-(t)\left|\kappa_l^-(t)\right\rangle = \pi_l^-(t)\left|\kappa_l^-(t)\right\rangle, \qquad (42)$$

where $\left|\kappa_l^-\right\rangle$ are sorted in the descending order of $\pi_l^-$. We remember the times $t_k^{\rm out}$ when the lowest eigenvalue $\pi_k^-(t_k^{\rm out})$ falls below the significance treshold, $\pi_k^-(t_k^{\rm out})/\pi_1^-(t_1^{\rm out}) < r_{\rm cut}$. Each time the corresponding mode $\kappa_k^-(t_k^{\rm out})$ is remembered as the outgoing mode, $\kappa_k^{\rm out} \equiv \kappa_k^-(t_k^{\rm out})$. The diagonalizing unitary matrix $U_k$ is also remembered and *applied to the remaining modes* since it corresponds to the frame rotation eq. (14). Immediately after that the mode $\kappa_k^{\rm out}$ is excluded from the propagation: the dimensions of $\widehat{\rho}_-(t)$ decrease by 1 at each time moment $t_k^{\rm out}$. In other words, we project $\widehat{\rho}_-(t)$ to the subspace of relevant modes of dimension $m_{\rm in}(t) - m_{\rm out}(t)$. Recurrently repeating this step we find the outgoing modes satisfying (O.1), (O.2), and (O.3).

### 4. The model parameters

The coupling coefficient $\chi_l^{\rm rel}(t)$ is found as

$$\chi_l^{\rm rel}(t) = \left[\widehat{a}_0(t), \widehat{\kappa}_l^{\rm rel\dagger}\right]_\xi$$
$$= \left\langle \phi(t)\left|\kappa_l^{\rm rel}\right\rangle. \quad (43)$$

The generator $D(t)$ is found as

$$D(t) = i\ln U_k/\left(t_k^{\rm out} - t_k^*\right), \qquad (44)$$

where $t_k^*$ is the time of the previous coupling/decoupling event.

Now we write the Schrodinger equation (15)-(16) in a Fock space which is truncated in maximal occupation of relevant modes, and solve the interaction quench problem.

## VI. CONCLUSION

In this work we propose the concept of bandlimited quantum noise. On a spin-boson model we demonstrate that the bandlimited quantum noise is effectively a discrete time process, which yields simple and intuitive equations of motion. These equations of motion are ready to be applied to specific problems. A numerical procedure to find the necessary coupling constants for a given spectral density $J(\omega)$ is presented.

The merit of the proposed approach is that it exploits the balance of complexity which is characteristic of the quantum dissipative motion [43]: only a bounded number of degrees of freedom are coupled to OQS at any given instant of time. Moreover, the population of these coupled degrees of freedom is bounded. This yields a microscopically-derived matrix product state (MPS) for the quantum dissipative evolution, with a bounded bond dimension. Interesting enough, we do not use SVD anywhere. In this work we provide an alternative way of how to obtain the MPS structures.

In this work we also propose a Monte Carlo simulation technique which samples the pure-state evolution interrupted by random quantum jumps. This has a benefit with respect to the density matrix methods due to a quadratically smaller dimension of the state vector. Importantly, these quantum jumps are irreversible and completed. There is no need to simulate the reversed quantum jumps [46, 47].

The model of bandlimited quantum noise also yields some conceptual insights. In particular, it shows that the classical records are stored in the asymptotically invariant entanglement structures. Given a pure state, we can compute the entropy of the ensemble of classical records which this pure state contains. It also presents a novel discrete-time input-output formalism.

The concept of bandlimited quantum noise is expected to be applicable to any kind of open system and environment. The work is in progress to cover more situations.

### ACKNOWLEDGMENTS

The work was carried out in the framework of the Roadmap for Quantum computing in Russia

## Appendix A: CLASSICAL KOTELNIKOV SAMPLING: VARIATIONAL FORMULATION

Here we reformulate the classical Kotelnikov sampling theorem so that it could be easily generalized to the case of quantum noise. We begin by noting that the ensemble of random classical signal samples $z(t)$ on a finite time interval $[0, T]$ is analogous to a many body state since it generates an infinite hierarchy of correlators

$$C_{\mathrm{cl}}\left(t_1, t_2, \ldots t_p \,|\, t_1', t_2', \ldots t_q'\right)$$
$$= \overline{z^*(t_1)\, z^*(t_2) \ldots z^*(t_p)\, z(t_1')\, z(t_2') \ldots z(t_q')}. \quad (A1)$$

Then the time correlator $C_{\mathrm{cl}}\left(t\,|\,t'\right) = \overline{z^*(t)\, z(t')}$ can be interpreted as a one-body reduced density matrix $\widehat{\rho}_{\mathrm{cl}}$ with matrix elements

$$\langle t |\widehat{\rho}_{\mathrm{cl}}| t'\rangle = \overline{z^*(t)\, z(t')}. \quad (A2)$$

Suppose we are given some normalized temporal function $\phi(t)$ denoted as $|\phi\rangle$. We may consider it as a DoF and ask a question: is it significant for $z(t)$ or we may discard it? In other words, we seek for a minimal number $m$ of DoFs $|\kappa_k\rangle$, $k = 1 \ldots m$, which play the most significant role in the representation of the signal, so that it effectively becomes $m$-dimensional, $z(t) \approx \sum_{k=1}^{m} z_m \langle t | \kappa_k\rangle$. To answer this question it is enough to consider the variance

$$\mathcal{K}[\phi] = \langle \phi |\widehat{\rho}_{\mathrm{cl}}| \phi\rangle. \quad (A3)$$

If $\overline{z(t)} = 0$, then a negligible $\mathcal{K}[\phi]$ indicates that the signal is effectively zero along the direction of $|\phi\rangle$, and it is irrelevant for the representation of $z(t)$. Then the most significant DoF $|\kappa_1\rangle$ is found variationally

$$|\kappa_1\rangle = \operatorname*{argmax}_{\kappa_1} \frac{\langle \kappa_1 |\widehat{\rho}_{\mathrm{cl}}| \kappa_1\rangle}{\langle \kappa_1 | \kappa_1\rangle}. \quad (A4)$$

Next we find the DoF $|\kappa_2\rangle$ which is the most significant after $|\kappa_1\rangle$. It satisfies the same variational principle under the constraint $|\kappa_1\rangle \perp |\kappa_2\rangle$ since it should be a new independent DoF. Repeating this recursively, we recognize this as the Rayleigh-Ritz formulation of the eigenvalue problem,

$$\widehat{\rho}_{\mathrm{cl}} |\kappa_l\rangle = \pi_l |\kappa_l\rangle, \quad (A5)$$

where the eigenvectors $|\kappa_l\rangle$ are sorted in the descending order of $\pi_l$. By choosing some relative significance threshold $r_{\mathrm{cut}}$ (e.g. $r_{\mathrm{cut}} = 10^{-4} \ldots 10^{-7}$), we keep only the first

$m$ modes which surpass this treshold, $\pi_k/\pi_1 > r_{\mathrm{cut}}$ for $k = 1 \ldots m$. I

It can be shown by a numerical computation that indeed $|\kappa_l\rangle$ span the same subspace as the sinc basis in eq. (5), and $m = 2BT$. We call the modes $|\kappa_l\rangle$ surpassing the significance treshold the *Kotelnikov modes*.

We conclude that the message of the Kotelnikov sampling theorem is that a bandlimited signal has a finite number $m$ of statistically significant DoFs on a finite interval $T$, and $m$ scales as $\propto BT$. These DoFs are found as eigenvectors of the largest eigenvalues of the one body reduced density matrix.

### Appendix B: SOLUTION FOR $\widehat{\rho}_+(t)$

We interpret the equation (29) as a stochastic equation in the Stratonovich sense

$$id |\Psi_{\xi+}\rangle = \left\{ dz^* \widehat{a}_0(t) + dz \widehat{a}_0^\dagger(t) \right\} |\Psi_{\xi+}\rangle. \quad (B1)$$

where $dz$ is increment of complex white noise (Wiener increment). The corresponding Ito form is

$$d |\Psi_{\xi+}\rangle$$
$$= \left\{ -idz^* \widehat{a}_0(t) - idz \widehat{a}_0^\dagger(t) - \left( \widehat{a}_0^\dagger(t)\, \widehat{a}_0(t) + \frac{1}{2} \right) dt \right\} |\Psi_{\xi+}\rangle. \quad (B2)$$

The one-body density matrix is

$$\rho_{+kl}(t) = \overline{\left\langle \Psi_{\xi+}(t) \left| \widehat{a}_k^\dagger \widehat{a}_l \right| \Psi_{\xi+}(t) \right\rangle}, \quad (B3)$$

where $\overline{(\cdot)}$ is stochastic average (over noise $z$ realizations). Using the identity

$$\widehat{a}_0^\dagger(t) = \sum_j \widehat{a}_j^\dagger \phi_j(t), \quad (B4)$$

where $\phi_j(t)$ satisfies eq. (32), we apply the Ito formula for the increment $d\rho_{+kl}(t)$. After some algebra we obtain:

$$d\rho_{+kl}(t) = dt \phi_k^*(t)\, \phi_l(t). \quad (B5)$$

Therefore, we get the equation (31).